\documentclass[twocolumn]{aastex63}

\usepackage{hyperref}
\usepackage{verbatim}
\graphicspath{{./}{/figures}}
\accepted{2023 January 24}
\submitjournal{ApJ}
\usepackage{amsmath}
\usepackage{amsfonts}
\shorttitle{GW from Distant Clusters}
\shortauthors{}

\begin{document}

\title{Ultra-Low Frequency Gravitational Waves from Massive Clusters at $z\sim1$}

\correspondingauthor{R. W. Romani}
\email{rwr@astro.stanford.edu}

\author[0000-0003-4963-164X]{David K. Wendt}

\affil{Department  of  Physics,  Stanford  University,  Stanford,  CA, 94305, USA}

\author[0000-0001-6711-3286]{Roger W. Romani}
\affil{Department  of  Physics,  Stanford  University,  Stanford,  CA, 94305, USA}

\begin{abstract}
Recent progress with pulsar timing array experiments, especially from the NANOGrav collaboration indicate that we are on the cusp of detecting significant signals from the inspiral of super-massive black hole binaries (SMBHB). While recent analysis has focused on nearby galaxies as possible sources of the loudest signals, we show that mergers in identified clusters at $z\sim 1$ can have larger strain amplitudes. We make an estimate comparing the nearby 2MASS redshift survey galaxy sample with the more distant MaDCoWS cluster sample, showing that the latter might be expected to contribute more, and louder, gravitational wave events. Thus the first individual source detections may well be from ultra-massive BH in clusters at $z\sim 1$, rather than nearby galaxies.
\end{abstract}

\keywords{pulsars:  general — gravitational waves}

\section{Introduction} \label{sec:intro}

In a recent paper \citet{2021ApJ...914..121A} have estimated the potential pulsar timing array (PTA) gravitational wave (GW) signal from hypothetical supermassive black hole binaries (SMBHB) within $\sim 0.5$Gpc, drawn from the 2MASS Redshift Survey (2MRS) galaxies (see also \citet{2017NatAs...1..886M} for an earlier estimate of the signal from 2MRS sources). They conclude that limits on on the gravitational wave amplitude show that, for at least a few of these nearby galaxies, SMBHB inspiral from major mergers cannot be taking place at present. However this work and other recent studies suggest that as PTA sensitivity advances, signals from the loudest sources might be expected soon. In speculating about the origin of such strong sources, one might recall the brief puzzlement over the initial LIGO detection of GW150914 \citep[$\sim 36+20M_\odot$ merging at $\sim 400$\,Mpc, ][]{2016PhRvL.116x1103A} before the more widely awaited nearby neutron star-neutron star mergers. Of course, a few authors \citep[e.g.][]{2010ApJ...715L.138B} had already concluded that more massive mergers will likely dominate the early LIGO detections. This is because, as an amplitude detection, one measures an orientation- and polarization-averaged rms strain
\begin{equation}\label{eq:h}
    h = \left ( \frac{32}{5} \right)^{1/2} \frac{G M_c}{c^2 D(z)} \left ( \frac{\pi G M_c f (1+z)}{c^3} \right )^{2/3} \sim \frac{M_c^{5/3}}{D}
\end{equation}
where $M_c= (M_1 M_2)^{3/5}/(M_1+M_2)^{1/5} = M_T [q/(1+q)^2]^{3/5}$ is the `chirp mass', $M_T$ is the total SMBHB mass, $q$ is the mass ratio $M_1/M_2$, $f$ is the observed gravitational wave frequency (from a binary rest frame orbital period $P=2/f(1+z)$ and $D(z)$ is the luminosity distance at redshift $z$. For such an amplitude measurement the sample size grows as $\sim D^3$, so more distant (rarer) massive binaries can dominate over a nearby sample in detections of high S/N sources standing above the background. This is the case for LIGO and may also be the case for PTAs.

%While several simulation studies show that more distant sources should contribute to the gravitational wave spectrum \citep[e.g.\;a quasar-based population in][]{2022ApJ...924...93C}, 
Following \citet{2021ApJ...914..121A} it is useful to identify individual candidate GW source sites from existing catalogs of known sources. These may help in searching for EM counterparts, pinning down source distances, following population evolution, etc. In this study we use a simple estimate of the catalog sources' potential strain amplitudes and merger rates to compare the nearby 2MRS sample with a high $z$ selection of massive cluster galaxies, drawn from the MaDCoWS \citep{2019ApJS..240...33G} analysis of the WISE IR survey data. Since the MaDCoWS clusters have a typical $0.8<z<1.3$, the WISE band 1 and band 2 images generally bracket the rest frame $K$ band, easing comparison of the two samples.

In \S 3 we compare the results for the two populations showing the (in retrospect unsurprising) result that the high $z$ cluster sample is fertile source of identifiable GW events, with sources standing out above the continuum generated by nearby (e.g. 2MRS) galaxies. Of course, without a complete sample at all $z$ and with no detailed treatment of the galaxy evolution, we will not have identified {\it all} of the loudest sources. But this simple analysis does emphasize that distant, relatively poorly studied hosts may be prominent in the early PTA detection set. Limitations of the analysis and implications of these results are discussed briefly in the conclusion. 

\section{GW amplitude/event rate estimates }\label{method}

Our basic approach is to assume a bulge-BH mass correlation, assume that the near-IR magnitudes of these late-type galaxies are bulge dominated and use these to estimate the total SMBH mass. This allows a scaled estimate of the characteristic GW signal amplitude for each source $h_0$, for a fiducial PTA observation frequency $f_{\rm obs}$. We then make a rough estimate of the major merger rate for each source, where we use the surveys to estimate the density of comparable luminosity companions in the source potential well. We define major merger companions as those having a bulge luminosity (and hence BH mass) $q>0.2\times$ that of the brighter primary. Combining the merger rate with the binary inspiral time at our fiducial GW frequency $f_{\rm obs}$ provides a probability that the pair (standing in for a similar pair drawn from the primary potential well) will currently be undergoing an active merger. Using these probabilities, we compare the GW $h_0$ distributions of the two populations, and identify the individual high $z$ clusters with relatively high probability of an active merger. 

Our approach is somewhat similar to that of \citet{2017NatAs...1..886M} but simpler, in that we do not use cosmological simulations or galaxy evolution to probe the details of merger rates and binary stalling. Instead we compute a simple merger rate and spectrum based on an analysis of catalog members, assuming that the central black holes of all mergers reach gravitational wave-dominated inspiral in less than a Hubble time. This analysis can be applied equally to low redshift (2MRS) and high redshift (MaDCoWS) source lists, to discuss their relative contribution to the gravitational wave spectrum. We start from near-all sky host samples of bulge-dominated galaxies, providing a list of the central black hole masses and candidate merger companions. 

\subsection{Parent Catalogs}

\begin{figure}[h!]
    \centering
    \vskip -8mm
    \includegraphics[scale=0.55]{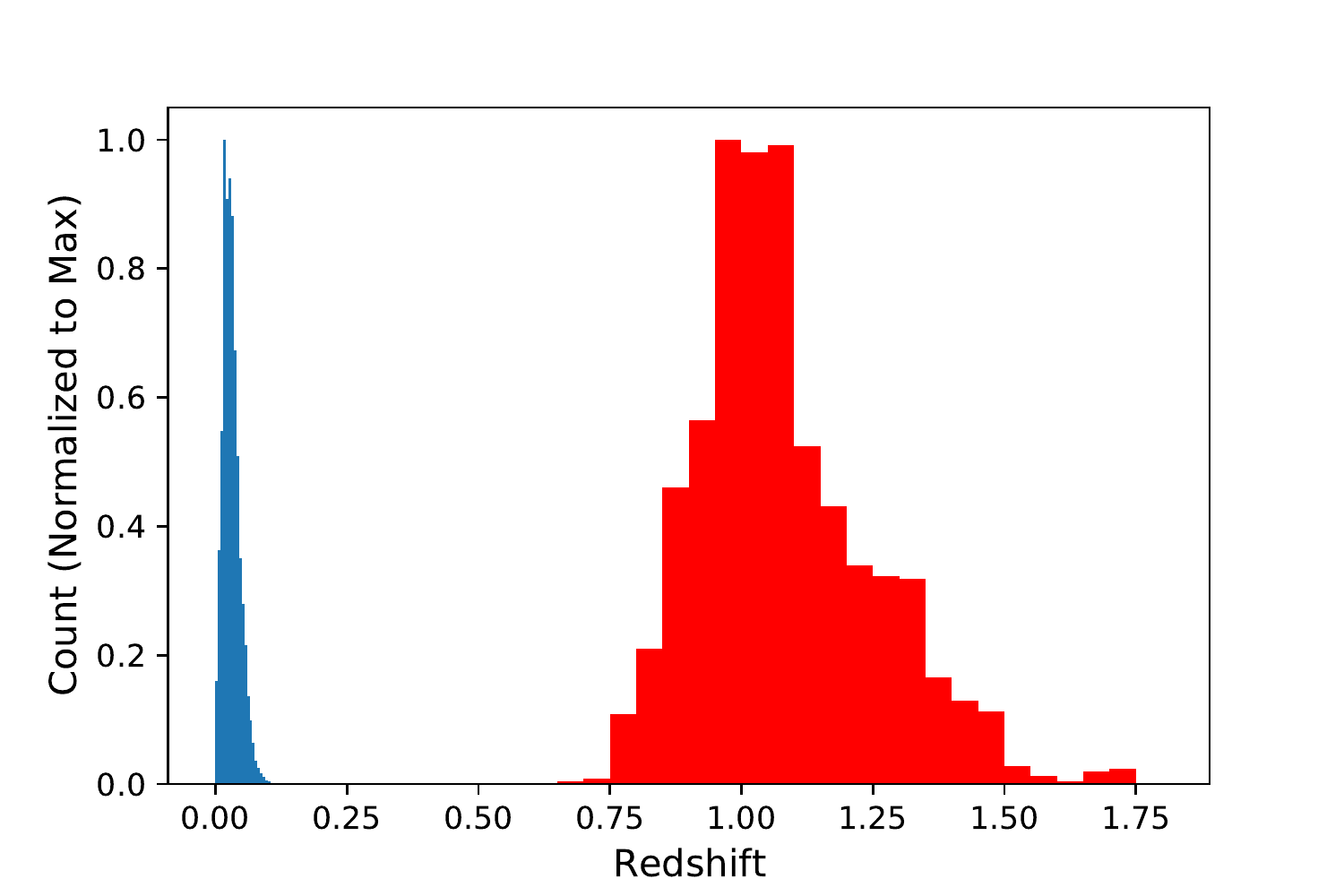}
    \caption{2MRS (blue) and MaDCoWS (red) redshift distributions. Note that 2MRS redshifts are spectroscopic, while most MaDCoWS values are photometric.}
    \label{fig:photo-z}
\end{figure}

For a catalog of low-redshift galaxies, we use the 2MASS Redshift Survey \citep[2MRS][]{2012ApJS..199...26H}. This contains extended sources from the Two Micron All-Sky Survey (2MASS) flagged as galaxies, with mean redshift $\langle z \rangle=0.03$ and the distribution shown in Fig.\,\ref{fig:photo-z}.

At higher redshift, the loudest signals will be from more massive galaxies (brighter bulges) and galaxies in dense, high interaction rate environments will be especially well represented. It is convenient to have a consistent set of redshifts for these sources. We therefore draw our high-z candidates from the Massive and Distant Clusters of WISE Survey (MaDCoWS), which identified clusters using infrared data from WISE and optical data from the PanSTARRS survey. The MaDCoWS selection is designed to identify clusters with $z>0.8$. Redshifts are estimated from photo-z analysis of the cluster members, augmented by a small number of spectroscopic measurements. The actual distribution peaks at $z\sim 1$ with a tail to $z\sim 1.5$ (Figure \ref{fig:photo-z}).

By restricting to clusters in the PanSTARRS footprint, i.e. the sky north of $-30^\circ$ and away from the galactic plane, MaDCoWS allows more detailed diagnostics of the cluster members and field galaxies. This sample covers 17,668 square degrees. 

Apparent magnitudes of candidate cluster members are extracted from the unWISE catalog \citep{2022RNAAS...6...62M}. We also use the PanSTARRS STRM \citep{2021MNRAS.500.1633B} catalog for source classification and photometric redshifts. 

\subsection{MaDCoWS Cluster Analysis}

\begin{figure}[h!]
    \centering
    \includegraphics[scale=0.45]{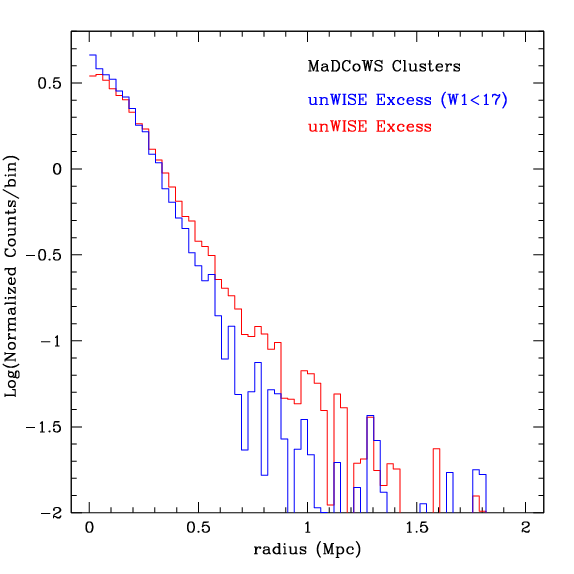}
    \caption{Background-subtracted excess density of unWISE source associated with MadCoWS clusters (normalized histograms of radial density distribution). The brightest sources are more centrally located.}
    \label{fig:clusters}
\end{figure}

We start by estimating the massive galaxy numbers in our cluster cores. UnWISE sources in $5^\prime$ fields around each cluster are identified with PanSTARRS STRM catalog sources, using a $3^{\prime\prime}$ matching radius. Sources flagged as stellar in the STRM catalog are excluded. Of 83,764 unWISE MaDCoWS cluster sources in the PanSTARRS overlap region, we find 18,660 with a STRM photo-z $z_p$ estimate. The remainder are unmatched (typically faint unWISE sources) or have no $z_p$. To exclude galaxies foreground to the MaDCoWS clusters, we cut all sources with well measured (photo-z error $\sigma_{z_p}< 0.1$) photo-z values $z_p<0.7$. This cut removes 30.6\% of the matched sources with $z_p$ (5,719 sources), mostly bright foreground galaxies. With the many faint unWISE sources lacking $z_p$, this cut removes only 6.8\% of all detections.

There will of course be additional faint foreground interlopers that cannot be be excluded via $z_p$, but we can statistically correct for these. We start by forming an estimate of the foreground $W_1$ (3.4$\mu$) apparent magnitude distribution, taking all objects passing the STRM non-stellar and $z_p$ cuts that lie $3^\prime - 5^\prime$ from the cluster centers. The interloper apparent magnitude distribution is formed by averaging over the annuli about all clusters, but we rescale the total areal density for each cluster to match to the actual density in that individual cluster's annulus. This rescaling provides some immunity to foreground variations due to intervening large scale structure and varying survey sensitivity. These scaled foregrounds are subtracted form the apparent magnitude histograms for each cluster core, giving a count of bright cluster members in a set of apparent $W1$ magnitude bins. 

Figure \ref{fig:clusters} shows the radial distribution of the foreground-subtracted $W1$ member counts in all of our clusters, using the cluster photo-z estimate $z_c$ to convert projected angle from the cluster center to radius. Note that these luminous galaxies are very strongly concentrated to the cluster cores, with the brightest ($W1<17$)  showing the highest concentration. The counts return to background levels by $\sim 2$Mpc, well within the $3^\prime - 5^\prime$ foreground annulus. Of course, fluctuations mean that for some clusters some W1 apparent magnitude bins will be negative after foreground subtraction -- we carry along these deficits in our final GW event rate sums so as to not overestimate cluster membership. A similar exercise gives the $W_2$ ($4.6\mu$) band cluster counts.

We focus on the brightest, most massive cluster members. For the primaries in major (mass ratio $>1/5$) mergers, we select $W1<17$ unWISE sources within the 0.5Mpc cluster core radius. At $z=1$ this magnitude cut-off indicates a typical central black hole of $M_{\rm BH} > 2.8 \times 10^9 M_\odot$ for standard bulge-mass relations (see below). To become this luminous ($>40\times L^\ast_K$) the hosts themselves are likely the product of many major mergers and in the dense cluster core environment such merging activity should be on-going.

We next estimate the rest $K$ band absolute magnitude of these MaDCoWS cluster members. The WISE colors of these sources are flat, so we convert catalog ($W_1$, $W_2$) pairs to absolute magnitudes $M_1$ and $M_2$ using $z_c$ in a flat $\lambda$-CDM concordance cosmology and then linearly interpolate to estimate the rest K-band ($2.2\mu)$ host magnitude as $M_K \approx M_2+(M_1-M_2)[3.83-1.83(1+z_c)]$. Applying this procedure to the background-subtracted apparent magnitude histogram yields a background-subtracted absolute $K$-band magnitude histogram for each cluster, with an occupation number for each magnitude bin. 

%\begin{figure}[h!]
%    \centering
%    \includegraphics[scale=0.25]{MC_fig3.png}
%    \caption{MaDCoWS cluster example (MOO J1234+4021). Left: WISE W1 band. Right: PanSTARRS optical. }
%    \label{fig:clusterim}
%\end{figure}

\subsection{2MRS Cluster Finding}

\begin{figure}[h!]
    \centering
    \vskip -0.3truecm
    \includegraphics[scale=0.6]{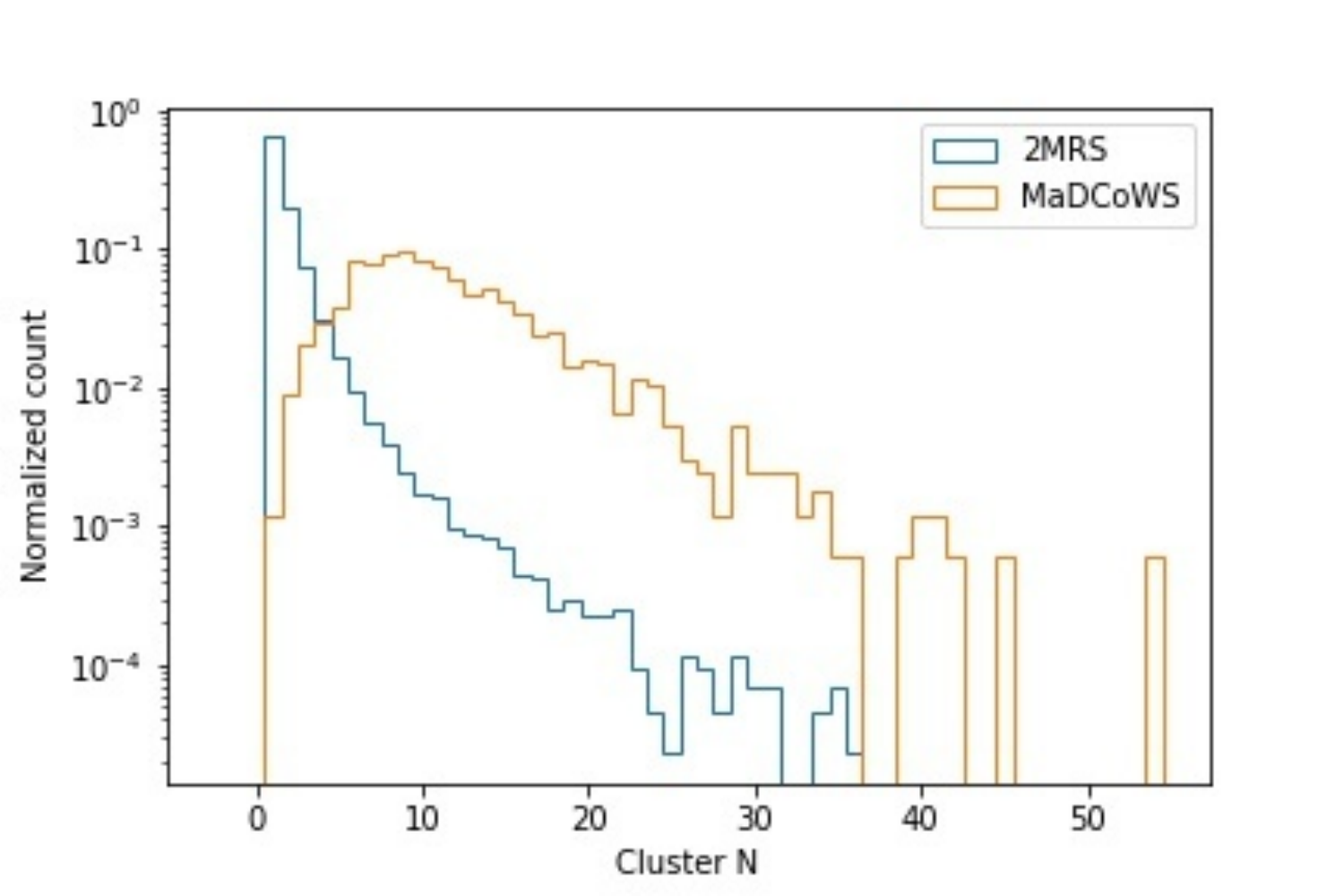}
    \vskip -0.3truecm
    \caption{Histogram of cluster galaxy counts within 0.5 Mpc. 2MRS in blue, MaDCoWS in orange.}
    \vskip -0.3truecm
    \label{fig:clcount}
\end{figure}

To treat the 2MRS galaxies on an equivalent footing, we have searched each galaxy for companions which might (in the future) produce major mergers. As for the MaDCoWS clusters, this analysis give a reasonable estimate for a steady state merger rate and hence gravitational wave chirp event rate. For each galaxy, we select all other 2MRS galaxies within an angular distance corresponding to 0.5 Mpc at the original source's redshift. If such a source's redshift is within two standard deviations of the original 2MRS target we assume that they are proximate and can contribute to the merger rate. As expected, most (67.6\%) 2MRS galaxies are isolates (see Figure \ref{fig:clcount}) so we expect a low merger rate compared to the rich clusters. However, the more luminous 2MRS objects undoubtedly have had several major mergers in the past. We have made no effort to select disturbed galaxies that suggest a recent merger. Instead we can optionally allow a fraction of the `singles' to have double cores with an ongoing GW-driven inspiral. Observationally, double cores are quite rare so this fraction should be modest. %Computationally we assign a low (e.g.\,10\%) probability to such on-going activity and treat the event as having a total BH mass corresponding to the galaxy bulge luminosity. 

\subsection{Strain Amplitude Estimation}

We now have a rest-frame K-band absolute magnitude $M_K$, for each galaxy and its candidate merger companions in the 2MRS analysis. Equivalently we have the individual clusters' absolute magnitude K-band histogram in the MaDCoWS analysis (with an occupation number for each bin). For the associated black hole masses there are many formulations of the well-known $M_\text{BH}$-$L_\text{bulge}$ relation; here we use the expression in \citet{10.1111/j.1365-2966.2007.11950.x} for the mass of the dominant BH and its putative companion
\begin{equation}
\log (M_\text{BH} / M_\odot) = 
- 0.37 M_K - 0.59,
\end{equation}
i.e. our rest frame $K=17$ corresponds to a BH mass of $2.8 \times 10^9 M_\odot$ at $z=1$. Thus for each proximate pair of 2MRS galaxies, or for each pair of MaDCoWS cluster luminosity bins (with occupation number) we have estimates for the constituent BH masses $M_{\text{BH},1},M_{\text{BH},2}$. Taking those pairs corresponding to a major merger $0.2 \leq M_2/M_1 < 1$, we calculate the strain amplitude with Eq. \eqref{eq:h}, assuming an observed gravitational wave frequency of $f_{\rm obs} = 1/5 \text{yr}^{-1}$, near the most sensitive frequency of modern PTAs. \

\subsection{Duty Cycle Estimation} \label{measurement-probabilities}

We wish to estimate what fraction of time a given galaxy pair (or bin pair for MaDCoWS) is `on' at an observed frequency $f_{\rm obs}$ in a post-merger GW-driven inspiral. Although massive galaxies are believed to have experienced many major mergers over a Hubble time \citep[e.g.][for a recent analysis]{2022ApJ...931...34F}  , there are few multi-core galaxies today. As long as one is not ejected in the merger, this means that the two BH must rapidly approach (through as yet incompletely modeled mechanisms) and coalesce once gravitational radiation losses take over. Thus the probability that a source is presently emitting at frequency $f_{\rm obs}$ becomes  
\begin{equation}
p = r_\text{merge} \tau,
\end{equation}
where, using the quadrupole formula for circular orbits $e=0$, we can write the lifetime at intrinsic frequency 
$f_\text{int}=f_\text{obs} (1+z)$ as  
\begin{equation}
\tau =E/{\dot E} = \frac{5c^5}{64(GM_\text{ch})^{5/3} (\pi f_\text{int})^{8/3}}
\end{equation}
with $M_{\rm ch}(M_1,M_2)$ for the pair given above. Note again that we implicitly assume that the post-merger evolution to the GW-dominated regime occurs in well under a Hubble time.

We now need a simple estimate for the galaxy merger rate in clusters of $N$ galaxies of mean mass $M_\text{gal}$ in a radius $R_\text{cl}$. These galaxies will have a typical velocity dispersion
\begin{equation}
v_\text{gal} \approx \sqrt{\frac{G\,N\,M_\text{gal}}{R_\text{cl}}}.
\end{equation}
We will assume a major merger occurs when a pair approaches within a characteristic galaxy separation $R_\text{gal}$. This gives a merger cross section $\pi R_\text{gal}^2 [1+(v_{\rm esc}/v_\text{gal})^2]$, where the second term accounts for gravitational focusing, via the typical galaxy escape velocity $v_{\text esc}^2 = 2GM_</R_\text{gal}$, where $M_<$ is the mass inside $R_\text{gal}$. If we assume that the galaxies are dominated by dark matter halos with flat rotation curves extending to the mean separation between cluster members $R_\text{cl}/N^{1/3}$, then $M_\text{gal} \approx R_\text{cl} M_</(N^{1/3}R_\text{gal})$. With $M_\text{cl} = N\,M_\text{gal}$ we get $(v_{\rm esc}/v_\text{gal})^2 = 2/N^{2/3}$. With these definitions,  the expected rate $r_\text{merge}$ for mergers between a given primary and any other galaxy in the bound system is
\begin{equation}
r_\text{merge} \approx \pi R_\text{gal}^2\left ( 1+\frac{2}{N^{2/3}} \right ) \cdot v_\text{gal} \cdot \left( \frac{N - 1}{\frac 43 \pi R_\text{cl}^3}\right);
\end{equation}
gravitational focusing is negligible except for poor clusters and groups.

In practice, a few physical scales are needed to use these estimates. For our massive field ellipticals, we have total (dark matter-dominated) mass $M_{\rm gal}= 10^{12}M_\odot$. The closest approach distance required for a major merger is uncertain; in the literature the core separation is often quite large, here we assume merger when the cores approach within $R_{\rm gal}=80$\,kpc. Similar values are for example used in modeling of the Milkyway-M31 merger \citep[e.g.][]{2020A&A...642A..30S}. The MaDCoWS cluster galaxies are on average $\sim 2.75$\,mag brighter and thus more massive, and larger \citep[with $R\sim M^{1/2}$][]{2015MmSAI..86..162C}. With $10\times$ larger mass, we adopt $R_{\rm gal}=240$\,kpc to initiate merger in the clusters. To obtain a typical cluster velocity dispersion $v_{\rm gal}\approx 1500 {\rm km\,s^{-1}}$, the $N\sim 10$ MaDCoWS galaxies should each represent a total mass of $M_{\rm gal}= 3\times 10^{13}M_\odot$ including dark matter and satellites. 

For 2MRS galaxy pairs, we assume that they are bound and compute the active merger probability from Equations 3-6 using $N=2$. The majority of 2MRS galaxies are however single, and would thus contribute nothing to the major merger rate. As above we might assume that some small fraction $f_\text{PM}$ of the 2MRS galaxies are immediate post-merger products, hosting an in-spiraling binary at their core. Surveys indicate that the fraction of galaxies presently in the post-merger state is small, so $f_\text{PM}\approx 0.1$ might be taken as an upper bound on this contribution. This is computed by setting $N=2$ for all parent 2MRS galaxies and multiplying the resultant probabilities by $f_\text{PM}$. This adds only a small 0.16 fraction to the total merger probability.
 
Each cluster galaxy pair with a given primary will have a different chirp mass depending on the secondary, so while we compute the orbital velocity (Equation 5) for all $N$ cluster members, we assign $N = 2$ in the $r_\text{merge}$ expression (Equation 6) for just that pair and compute the corresponding $\tau$ using the $M_\text{ch}$. For 2MRS, this results in a single value of $h$ and of $p$ for each pair of galaxies, while for MaDCoWS clusters we have such values for every pair of bins. For the clusters we multiply the active probability from Eq. 3 by the bin occupation numbers to find the total probability for that bin pair. As noted above, due to foreground subtraction, some MaDCoWS mass bins will have negative occupation numbers and when either the primary or the companion histogram bin is negative, we subtract the corresponding amplitude in our summed probability-weighted $h$ distribution.

\section{Results and Conclusions}\label{results}

\begin{figure}[h!]
    \centering
    \vskip -1.4truecm
    \hskip -12.1mm
    \includegraphics[scale=0.45]{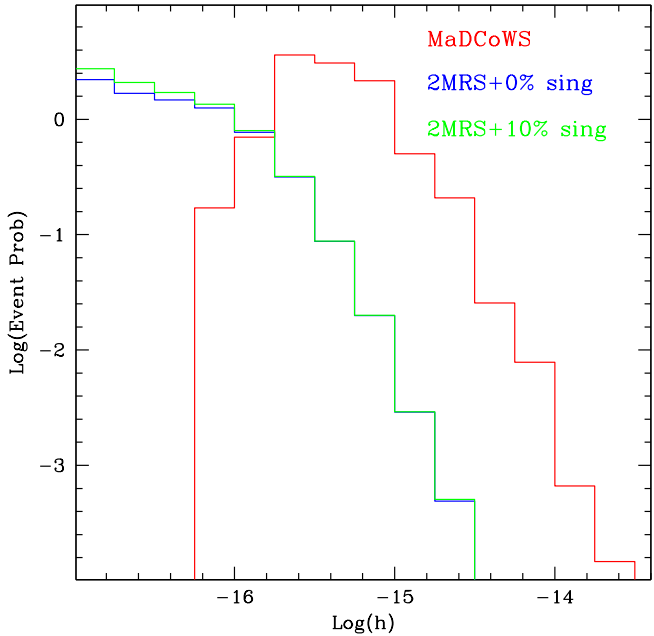}
    \vskip -2.4truecm
    \caption{The total probability-weighted histograms of strain amplitude are plotted for 2MRS (blue), 2MRS with 10\% of single galaxies taken to have post-merger double cores (green), and MaDCoWS (red).}
    \label{fig:h0}
\end{figure}

Figure \ref{fig:h0} shows the summed probability, at observed frequency $f_{\rm obs} = 1/5 \text{yr}^{-1}$, for the 2MRS and MaDCoWS components at each observed strain amplitude $h$. Several aspects are worth comment. First, even after foreground subtraction, the typical MaDCoWS galaxy has 10-15 companions and in the dense cluster cores the collision (i.e. merger) rate is large. Of course the BH masses of these extremely bright ellipticals are also large and so the inspiral lifetimes at our selected $f_\text{obs}$ are short. These effects combine to make MaDCoWS the dominant component at large $h$ despite the order of magnitude larger distance, while the larger local 2MRS population contributes a small $h$ background to these events. The crossing point for the intensity from these two catalogs is at ${\rm log}(h)\approx -16$. The 2MRS contribution peaks at lower $h$ as expected for the smaller bulge and hole masses. 

Recall that we have assumed evolution to gravitational wave domination in less than a Hubble time. However, in some cases ($>10^9 M_\odot$ e=0 binaries undergoing dry, gas-free mergers in core-depleted ellipticals) the merger may stall for several Gy at sub-pc scales \citep{2015MNRAS.454L..66S}. This could lead to an unresolved population of core binaries which would result in delayed gravitational wave signals from merged cluster galaxies. In fact, if such dry stellar dynamics-driven mergers dominate the production of the most massive ellipticals, then \citet{2023MNRAS.519.2083I} suggest that as many as 50\% of these galaxies may host unresolved SMBHB. While the merger events generating these binaries would be captured in our sums, the gravitational wave signal would be appear latter, e.g. by $\Delta z \approx -0.2$ for a Gyr delay in the typical MaDCoWS cluster.  

For the typical less massive 2MRS galaxies, events in presently isolated galaxies should contribute little. \citet{2023MNRAS.519.2083I} infer $<0.1$ hidden binary fraction in these galaxies, which we have seen produces  $<$16\% increase in the total event rate, almost all with ${\rm log}(h)<-16$. We will thus ignore this small contribution. To match more closely the real sky flux, we do correct for the limited sky coverage of our parent surveys. For 2MRS this means the probabilities are multiplied by $1.106$, for MaDCoWS by $2.335$. We conclude that very massive BH in $z\sim 1$ clusters can dominate the nearby field galaxy contribution in many GW frequency bins.

For an alternative presentation of these results, we can simulate realizations of the spectrum of the gravitational wave sky, following explicitly the contributions from individual 2MRS galaxies and MaDCoWS clusters. This is natural in our computation since each galaxy (or, for clusters, histogram bin) pair represents an object that spirals in through a range of $f_\text{obs}$. Since the strain amplitude scales as $h\sim f_\text{obs}^{2/3}$ while the lifetime (and hence on-sky probability) scales as $\sim f_\text{obs}^{-8/3}$, we can draw realizations of the sky at each observed frequency, using all pairs from the real objects in the surveys. Summing up these $h$ we obtain a realization of their contributions to the GW spectrum. 

\begin{figure}[h!]
    \centering
    \vskip -1.4truecm
    \hskip -12.1mm
    \includegraphics[scale=0.45]{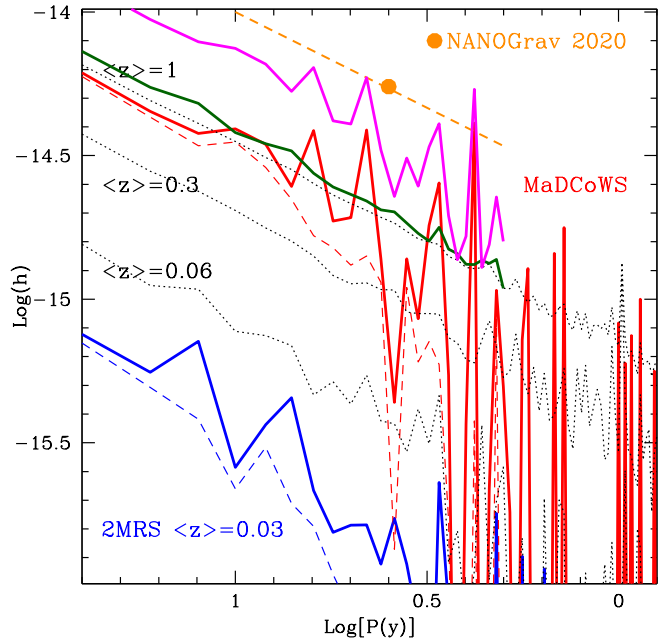}
    \vskip -2.7truecm
    \caption{A realization of the GW sky with contribution of 2MRS (blue) and MaDCoWS cluster (red) SMBHB. The dashed line removes the loudest source in each frequency bin. Note the clusters contribute occasional sources well above the `continuum'. The orange dot at top shows the GW amplitude spectrum inferred from excess noise in the current NANOGrav PTA analysis, with a dashed line for the continuum slope. Assuming that the local 2MRS population extends to $\langle z \rangle = 1$ we can compute a GW background. The thin dotted lines show the net (background) spectrum of such sources extending to increasing distance. Adding to this background the 2MRS and MaDCoWS spectra, we get total spectra (green and magenta, respectively). We see that only MaDCoWS provides SMBHB that, in a few frequency bins, may rise above the background confusion sufficiently for individual source detection.}
    \label{fig:GWspec}
\end{figure}

In Figure \ref{fig:GWspec} we show such a realization, computed in the frequency bins of a 30y PTA experiment. To show the contribution of the loudest GW source in each frequency bin, the dashed lines show the summed amplitude excluding this loudest source. As expected from Figure \ref{fig:h0}, the MaDCoWS sources dominate in the observable 1-10y band. More interestingly, rare, very massive pairs in these distant clusters often provide the largest individual sources in the PTA sensitive band. In our simulations we keep track of all sources that might be observable in the next decade of PTA experiments, selecting pairs that give $h>2\times 10^{-15}$ at $P_{\rm obs} < 10$y. 2MRS sources very seldom pass this threshold in our simulations, while the typical realization has several MaDCoWS galaxies causing events audible at this level. In the particular simulation plotted, for example, we have 
a $(11+11) \times 10^9M_\odot$ BH pair in cluster MOO JJ224$-$1514 at $z=1.40$ giving $(f_{\rm obs},{\rm log}h) = (0.80,-14.46)$, 
a $(11+7) \times 10^9M_\odot$ BH pair in cluster MOO J1444+0827 at $z=1.13$ giving $(0.68,-14.44)$,
a $(7+5) \times 10^9M_\odot$ BH pair in cluster MOO J0345$-$2913 at $z=1.08$ giving $(0.47,-14.61)$ and
a $(7+7) \times 10^9M_\odot$ BH pair in cluster MOO J1050+0947 at $z=1.08$ giving $(0.38,-14.39)$.

Of course there are many more distant 2MRS-like field galaxies contributing to the total gravitational wave background. We assume here that the population is like that of the $\langle z \rangle = 0.03$ sample, but at larger distance. In Figure \ref{fig:GWspec} we re-draw from the 2MRS sample, but at progressively larger distances, so that the number of draws scales as $d^3$ and amplitude of each draw scales as $1/d$. The faint dotted lines show the rms amplitude from the combined signal out to progressively larger redshift; $\sim 3 \times 10^4$ 2MRS-like volumes contribute out to $z\sim 1$. We can consider this combined 2MRS-like population to represent the background continuum of gravitational wave sources. The level is quite consistent with background level estimated in previous studies, e.g. Fig 1 of \citet{2015MNRAS.451.2417R}. Note that the background spectrum becomes less sparse at high frequencies when one includes the many distant sources. Note also that this `2MRS-like' sum slightly under-produces the total GW background which may have been detected by the NANOGrav experiment as excess noise at the $h\sim 1.97\times 10^{-15} f_{yr^{-1}}^{-2/3}$ level near their most sensitive range of $P_{\rm obs} \sim 4$\,y \citep{2020ApJ...905L..34A}. This supports our choice of $R_{\rm gal} = 80$\,kpc for the 2MRS merger distance as plausible: the background amplitude scales as $R_{\rm gal}^2$, so slightly large merger separations would saturate the present observational bounds.

We now ask how well the local 2MRS objects and the distant MaDCoWS mergers show up against this background. This is shown by the dark green and magneta spectra, respectively, focusing on gravitational wave frequencies $<1/(2{\rm y})$ where the background is well defined with low stochasticity. Note that in the green curve the brightest 2MRS sources (from the blue curve) are hardly visible against the total background. In contrast several MaDCoWS sources show up quite well, with power close to the present observational limits.
With this simple model it is evident that, when examining binaries expected from current source catalogs nearby 2MRS-type source will seldom stand out against the background noise, while at least a few frequency bins may be dominated by loud, distant cluster sources. 

The simple computations in this paper suffice to demonstrate how distant powerful sources may contribute to the early PTA individual source detections, just as they have for LIGO. We have made many simplifications that could be mitigated in further, more detailed analysis. First, a true full-sky distant cluster catalog (e.g. from eROSITA or CMB surveys) could provide a more complete finding list at these redshifts, (avoiding our extrapolation from the PanSTARRS and MaDCoWS footprints). Similarly, large area intermediate redshift samples may become available in the next decade allowing the massive bulge census to extend from $z\approx 0.1$ to $z > 0.8$. In modeling mergers, more detailed galaxy-hole mass relations might be used, possibly including host evolution to better exploit the W1 and W2 magnitudes. Also more detailed estimates of the maximum merger separation in cluster environments would certainly help. However, we suspect that none of these factors will alter our basic conclusions: that nearby 2MRS galaxies will only occasionally be the loudest sources to PTA experiments and that we will likely need to look far for the hosts of initial PTA gravitational wave signals, which may well be dominated by exceptionally massive galaxies in rich clusters. Deep surveys sensitive to $z>1$ will be needed to study possible electromagnetic counterparts, host clusters (and galaxies) and SMBHB evolution. 

\acknowledgements
We thank the second ananymous referee for a number of suggestions which helped improve the paper. We also thank the Stanford Physics Department for supporting DKW's research effort and the Center for Space Science and Astrophysics (CSSA) for supporting this publication.

\bibliographystyle{aasjournal}
\bibliography{mainpaper}
\end{document}